%% file: ICRC2025_template_IceCube.tex
\documentclass[a4paper,11pt]{article}

\usepackage{pos}
\usepackage{lipsum} 
\usepackage{subcaption}
\usepackage{amsmath}
\usepackage{wrapfig}
\usepackage{graphicx}

\title{Measurement of All Flavor PeV Neutrino Flux using Combined Datasets from IceCube}

\ShortTitle{Combined Measurement for PeV region}

\author{The IceCube Collaboration \\{\normalsize \normalfont(a complete list of authors can be found at the end of the proceedings)}\\}

\emailAdd{emre.yildizci@icecube.wisc.edu}
\emailAdd{zoe.rechav@icecube.wisc.edu}
\emailAdd{lu.lu@icecube.wisc.edu}

\abstract{

The IceCube Neutrino Observatory has detected astrophysical neutrinos with energies ranging from a few TeV to multiple PeV. To extend the energy range beyond 10 PeV, we combine throughgoing tracks from the northern sky, high-energy starting tracks, contained cascades and uncontained cascades from the entire sky. This extension is critical for testing models that predict a common origin for ultra-high-energy cosmic rays (UHECRs) and neutrinos. We present our analysis, designed to achieve the highest sensitivity to the PeV neutrino flux through advanced modeling of atmospheric backgrounds, refined ice systematics, and improved energy resolution in event reconstructions enabled by a combination of machine learning and likelihood-based techniques.

\vspace{4mm}

{\bfseries Corresponding authors:}
Emre Yildizci$^{1*}$, 
Zoe Rechav$^{1}$, 
Lu Lu$^{1}$\\
{$^{1}$ \itshape University of Wisconsin-Madison}\\[4mm]
$^*$ Presenter
}

\input{ICRCdetails.tex}

\begin{document}

\maketitle

\section{Introduction}\label{sec1}
The recent detection of an extremely high-energy neutrino candidate ($220\substack{+570 \\ -110}$ PeV) by KM3NeT \cite{KM3NeT:uhe-neutrino} raises intriguing questions about its origin—particularly given the absence of neutrino detections above 50 PeV by IceCube and the Pierre Auger Observatory. A precise characterization of the high-energy diffuse neutrino flux in IceCube is essential to better address this potential tension. Additionally, it will help constrain various generic source models that predict common origins for ultra-high-energy cosmic rays and neutrinos.

Since IceCube’s first discovery in 2013 \cite{Aartsen:evidence-2013}, the characterization of the high-energy astrophysical neutrino flux has been one of the experiment’s major focuses. Measurements using individual detection channels yielded largely compatible single power law (SPL) spectra \cite{estes, Abbasi:diffuse_numu}. Subsequent efforts combining different detection channels — namely, starting tracks and cascades within a dataset \cite{mese-icrc25}, and, following the pioneering multi-dataset combined analysis in \cite{Combined-fit-1}, throughgoing tracks with starting cascades across multiple datasets \cite{Abbasi:globalfit-2023} — reported substructures in the low-energy regime, though they still lacked the statistical power to resolve features above PeV energies.

Building upon this foundation, we present our analysis that incorporates both starting and throughgoing tracks, as well as a neural network–based cascade sample encompassing both contained and uncontained cascades, to enhance sensitivity at PeV energies. Furthermore, we update the event reconstructions, atmospheric neutrino and muon background models, and the ice modeling used in the detector response for each individual sample.

\section{Dataset description and updates}\label{sec2}
\begin{figure}[h!]
\centering
\vspace{-0.5pt}

\begin{subfigure}[b]{0.222\textwidth}
\centering
\includegraphics[width=\linewidth]{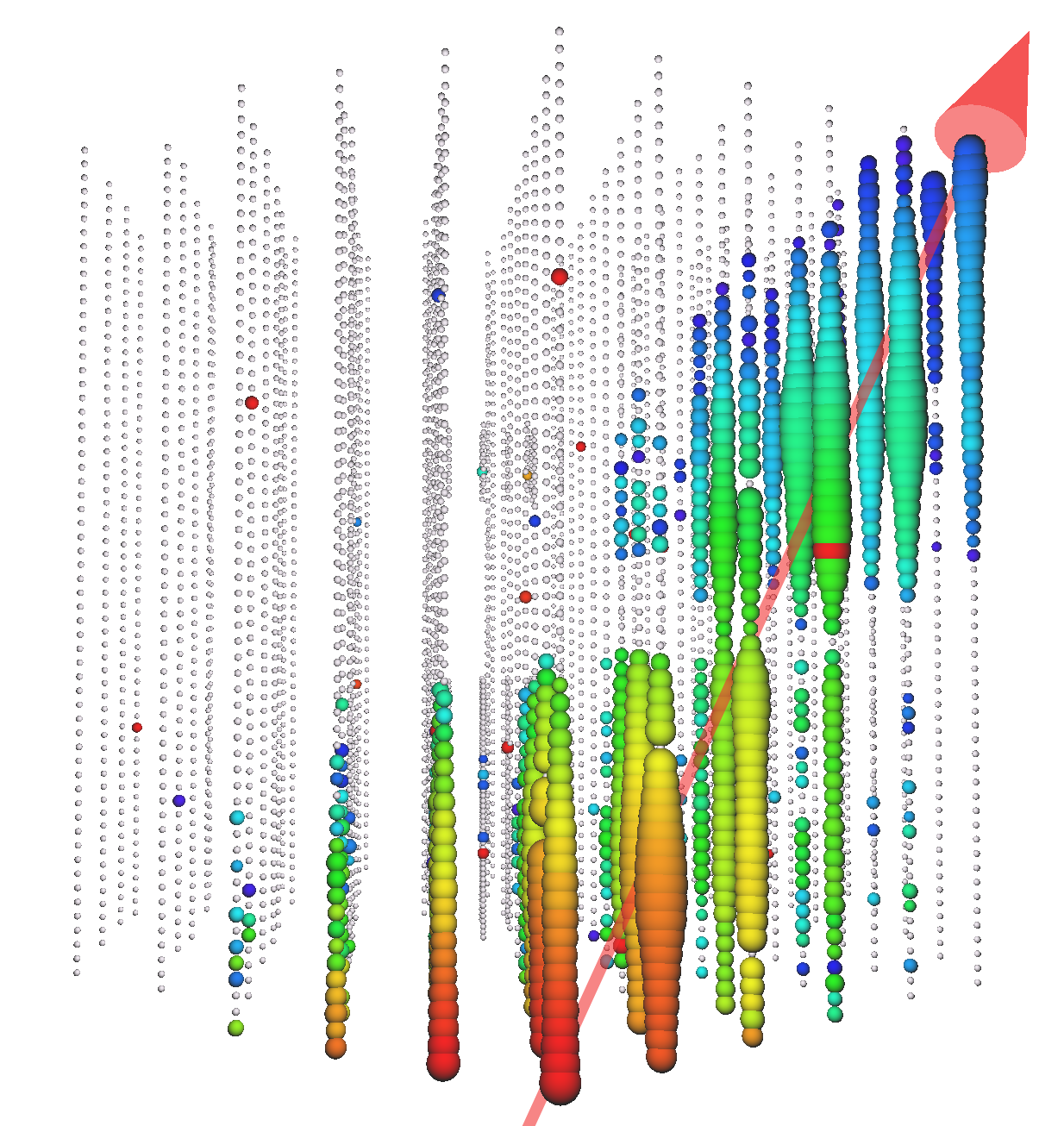}
\caption{Throughgoing track}
\label{fig:1}
\end{subfigure}
\hspace{0.01\textwidth}
\begin{subfigure}[b]{0.218\textwidth}
\centering
\includegraphics[width=\linewidth]{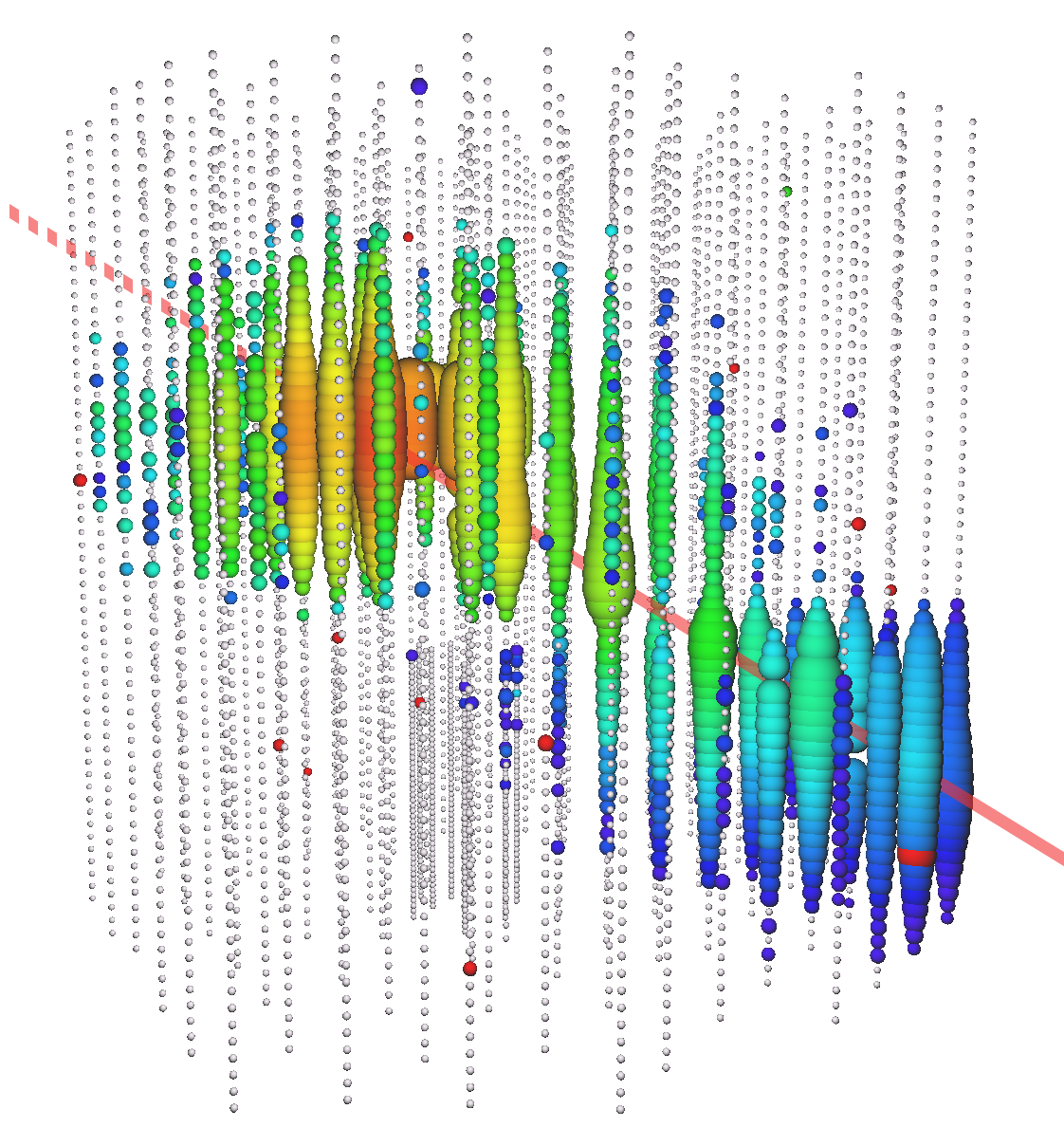}
\caption{Starting track}
\label{fig:2}
\end{subfigure}
\hspace{0.01\textwidth}
\begin{subfigure}[b]{0.224\textwidth}
\centering
\includegraphics[width=\linewidth]{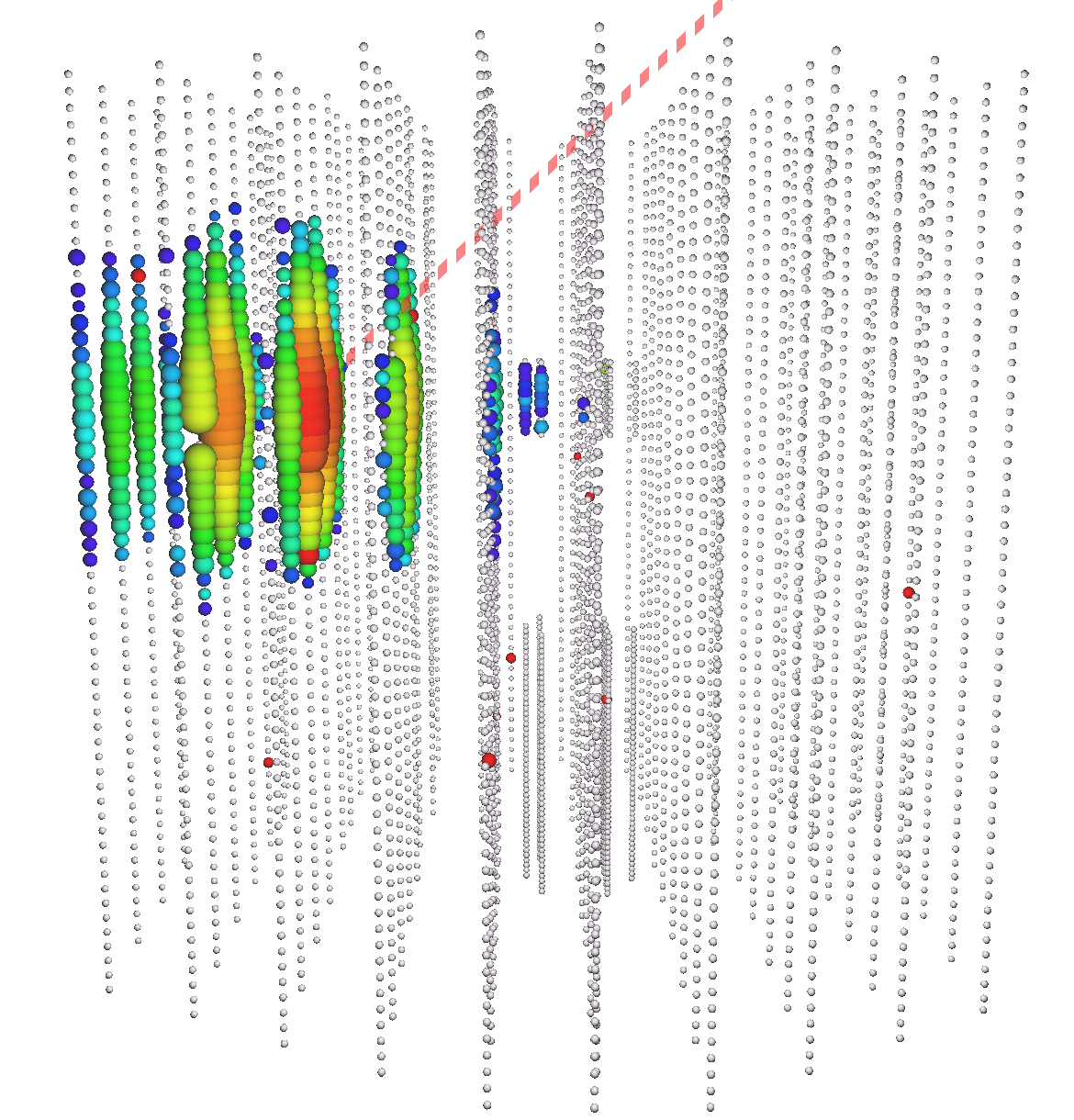}
\caption{Contained cascade}
\label{fig:3}
\end{subfigure}
\hspace{0.01\textwidth}
\begin{subfigure}[b]{0.218\textwidth}
\centering
\includegraphics[width=\linewidth]{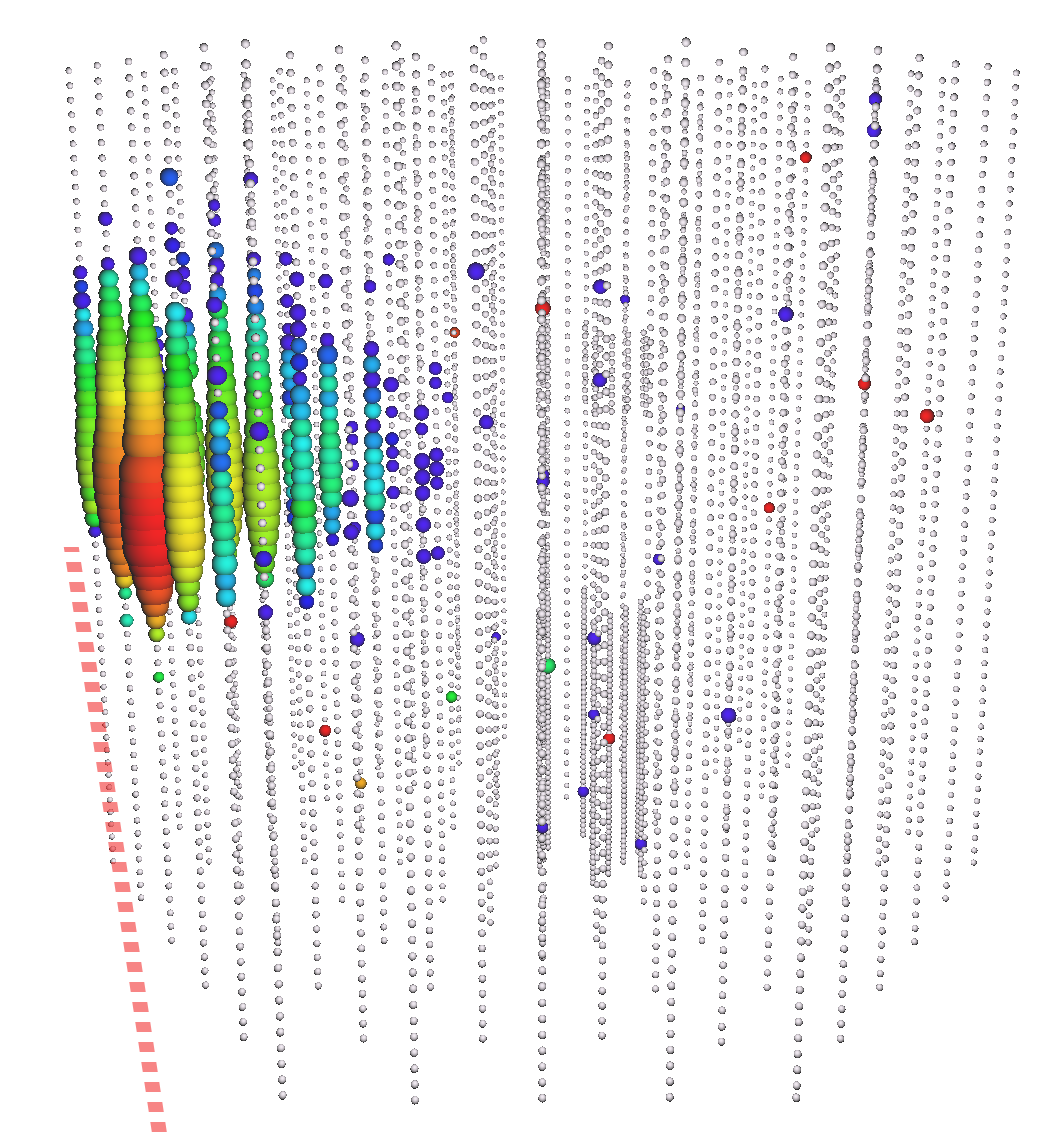}
\caption{Uncontained cascade}
\label{fig:4}
\end{subfigure}
\vspace{-2.5pt}

\caption{Example visualizations of simulated high-energy events for each of the four morphologies. Dashed red lines indicate the interacting neutrino, where applicable. The solid red line represents the throughgoing track produced by a neutrino interaction occurring outside the detector. Colors indicate the arrival time of the first detected light at the optical modules, with red corresponding to the earliest and blue to the latest.}
\label{fig:event-views}
\vspace{-5.5pt}
\end{figure}

To achieve the highest statistics of all-flavor neutrinos at extreme energies, we include multiple datasets targeting four event morphologies: throughgoing tracks, starting tracks, contained cascades, and uncontained cascades. Example visualizations of simulated high-energy events for each morphology are shown in Figure~\ref{fig:event-views}.

\subsection{Throughgoing track sample}\label{subsec1}
The throughgoing track sample used in this analysis consists of upgoing events from the Northern Hemisphere, which have long been utilized by IceCube, as the Earth naturally acts as a filter against the experiment’s most prominent background: atmospheric muons. To further suppress this background, the throughgoing track sample employs a boosted decision tree, achieving a neutrino purity of 99.8\% \cite{Abbasi:diffuse_numu}. With the atmospheric muon background effectively removed, the sample includes upgoing tracks which can originate from far outside the detector yielding a very large effective area for the event selection. The substantial event statistics in the sample also help constrain systematic uncertainties in background modeling and detector response.

\begin{figure}[h!]
\centering
\includegraphics[width=0.9\linewidth]{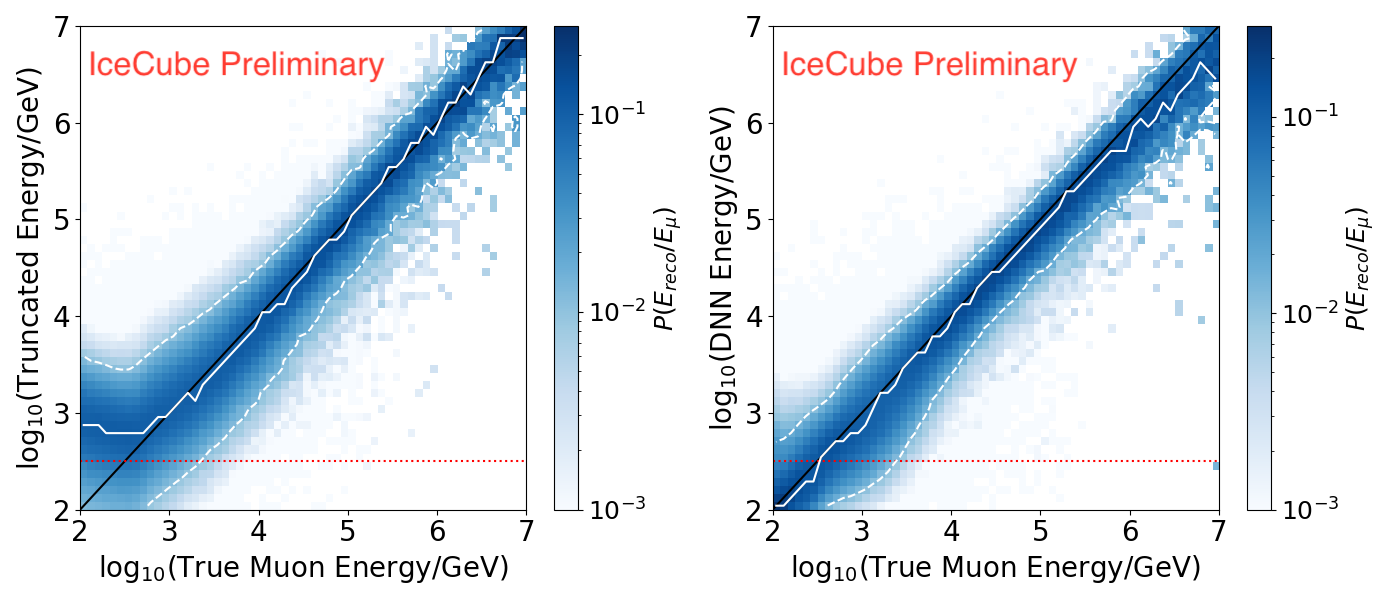}
\caption{Comparison of reconstructed and true muon energy with truncated energy (left) and DNN energy (right). The color bars show the probability of reconstructed energy, given true muon energy. Solid (dashed) white lines indicate the median (10\% and 90\% quantiles) of the distributions. Red dotted lines indicate the reconstructed energy binning threshold.}\label{fig:NT_reco_2D}
\vspace{-5pt}
\end{figure}

The reconstruction method previously used for these throughgoing muon tracks, known as truncated energy \cite{ABBASI:truncated}, estimates the average energy loss ($dE/dx$) by excluding segments with large stochastic energy depositions, providing a value proportional to the muon energy. However, this method becomes ineffective for low-energy muons, where $dE/dx$ is no longer proportional to the muon energy, leading to poor reconstruction performance. To address this, we replace the truncated energy method with a deep neural network (DNN) based energy reconstruction, DNN energy \cite{Abbasi:dnn-reco}, which mitigates the bias below $\sim$1 TeV and improves the energy resolution, as illustrated in Figure~\ref{fig:NT_reco_2D}.

\subsection{Starting track sample}\label{subsec2}
The full-sky starting track sample is created by applying a neural network–based event-type classifier \cite{dnn-classifier} to the High Energy Starting Events (HESE) sample \cite{hese-7.5yr}. The classifier serves two purposes: first, to select starting track events while leaving starting cascades to the cascades sample thereby minimizing overlap, and second, to further suppress the atmospheric muon background. This suppression occurs in addition to the traditional veto layer-based cuts in the HESE selection, by classifying atmospheric muons as throughgoing or stopping tracks. Furthermore, events that pass the HESE selection but are labeled as throughgoing or stopping tracks form the atmospheric muon–dominated “muon-tagged” sample. This control sample is used to constrain the normalizations of the atmospheric muon fluxes: conventional muons from pion and kaon decays, and prompt muons from charmed meson decays.

Additionally, in the Southern sky, the atmospheric neutrino background can be suppressed by the so-called self-veto effect \cite{self-veto}, in which accompanying muons from the same air shower deposit light in the outer veto layer as they enter the detector, causing the event to be rejected. This results in significantly higher astrophysical neutrino purity.

\begin{wrapfigure}{r}{0.55\textwidth}
\vspace{-7.5pt}
\includegraphics[width=\linewidth]{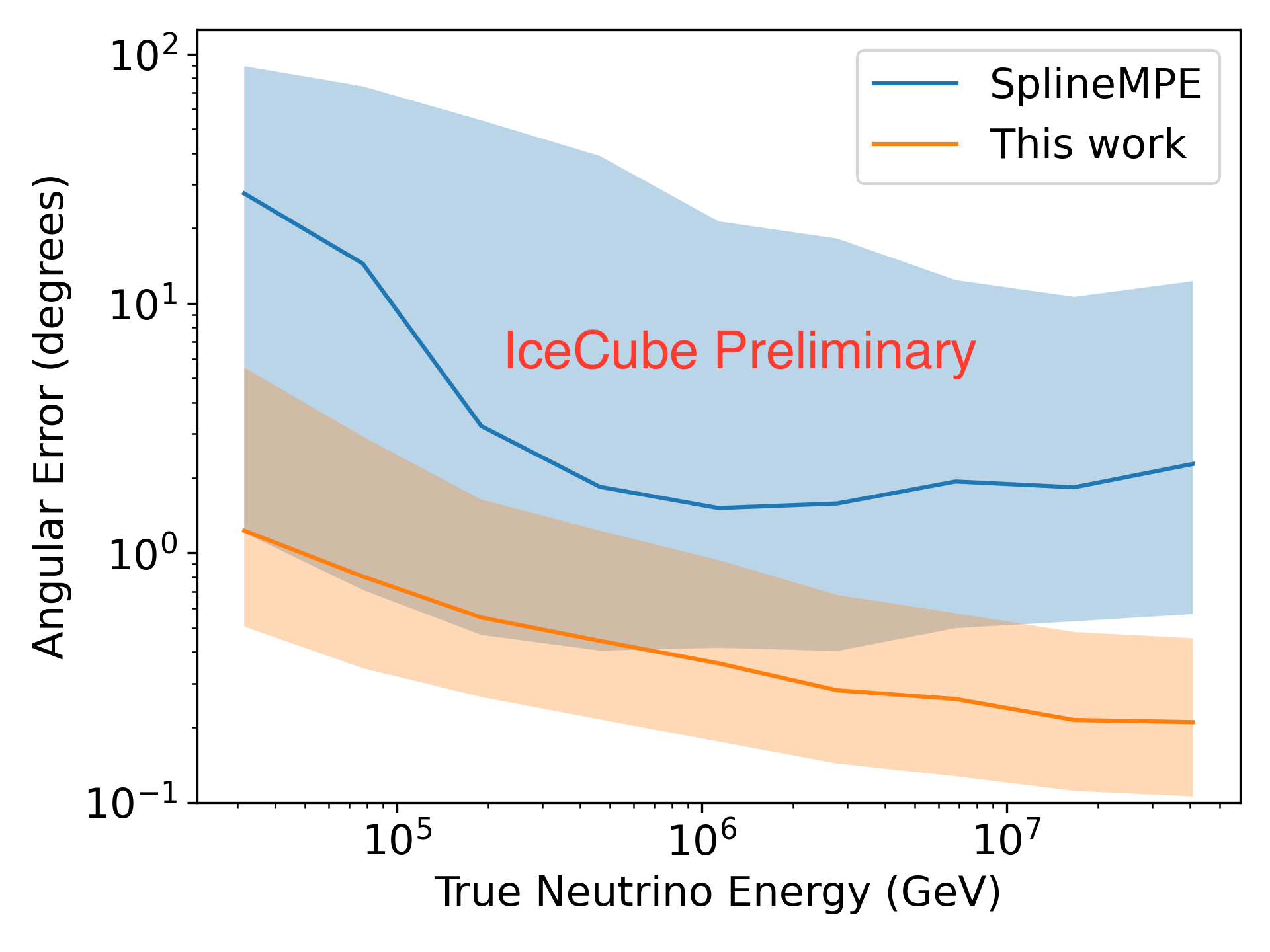}
\vspace{-20.5pt}
\caption{Comparison of angular errors between the standard track reconstruction (SplineMPE) and this work for starting track events. Solid lines represent the median and shaded area boundaries are the first and third quartiles.}\label{fig:hese-reco}
\vspace{-2.5pt}
\end{wrapfigure}

The sample is dominated by muon neutrino charge-current (CC) events in which energy depositions from both the hadronic shower and the muon track are visible in the detector. This enables the reconstruction of the neutrino energy, rather than the muon energy. However, the directional reconstruction is more challenging for starting tracks, as the track length inside the detector is typically shorter than for throughgoing events, providing less directional information. To address this, we developed a reconstruction method optimized for such events, including cascade-like events with limited visible track segments. This method adopts a hybrid approach, using the results of traditional track and cascade reconstructions as seeds for a final track reconstruction, in which energy depositions from the initial hadronic shower and stochastic losses from high-energy muons are approximated as cascade-like segments along the muon trajectory \cite{millipede}. As a result, the method achieves significantly improved angular resolution compared to the standard track reconstruction typically used for throughgoing events in IceCube \cite{Abbasi:splineMPE}, as shown in Figure~\ref{fig:hese-reco}.

\subsection{Contained and uncontained cascade sample}\label{subsec3}
The cascade sample used in this analysis is based on the dataset originally developed to establish the Galactic plane neutrino flux \cite{dnn-cascade}, with updated event reconstructions and self-veto modeling as described in \cite{zoe-dnn}. The contained cascades in this sample, which have been extensively used in IceCube’s studies of the astrophysical neutrino spectrum, including its initial discovery \cite{Aartsen:evidence-2013}, have statistics comparable to those in \cite{Abbasi:globalfit-2023}. In contrast, the uncontained cascades — most recently employed in the detection of a Glashow resonance event and the subsequent characterization of the PeV region by IceCube~\cite{GR} — make up two-thirds of the total cascade dataset above 100 TeV ~\cite{zoe-dnn} in this analysis and play a significant role in constraining the astrophysical flux at the highest energies.

\section{Background modelling updates}\label{sec3}
\subsection{Atmospheric neutrino treatment}\label{subsec4}

Historically, the uncertainties in the atmospheric neutrino flux were modeled using a scheme based on the Barr parametrization \cite{barr} for hadronic interaction uncertainties, applicable only to the conventional flux. In addition, an interpolation parameter between two primary cosmic ray models, a spectral shift parameter, and overall normalizations calculated using MCEq \cite{mceq} were used for both conventional and prompt fluxes. To achieve a more modern and comprehensive characterization of the atmospheric neutrino flux, we adopt the DAEMONFLUX scheme \cite{daemonflux}, which incorporates the Global Spline Fit (GSF) modeling of cosmic ray fluxes and a Data-Driven hadronic interaction Model (DDM).

\begin{figure}[h!]
    \begin{minipage}{0.48\linewidth}
        \raggedleft
        \includegraphics[width=0.97\linewidth]{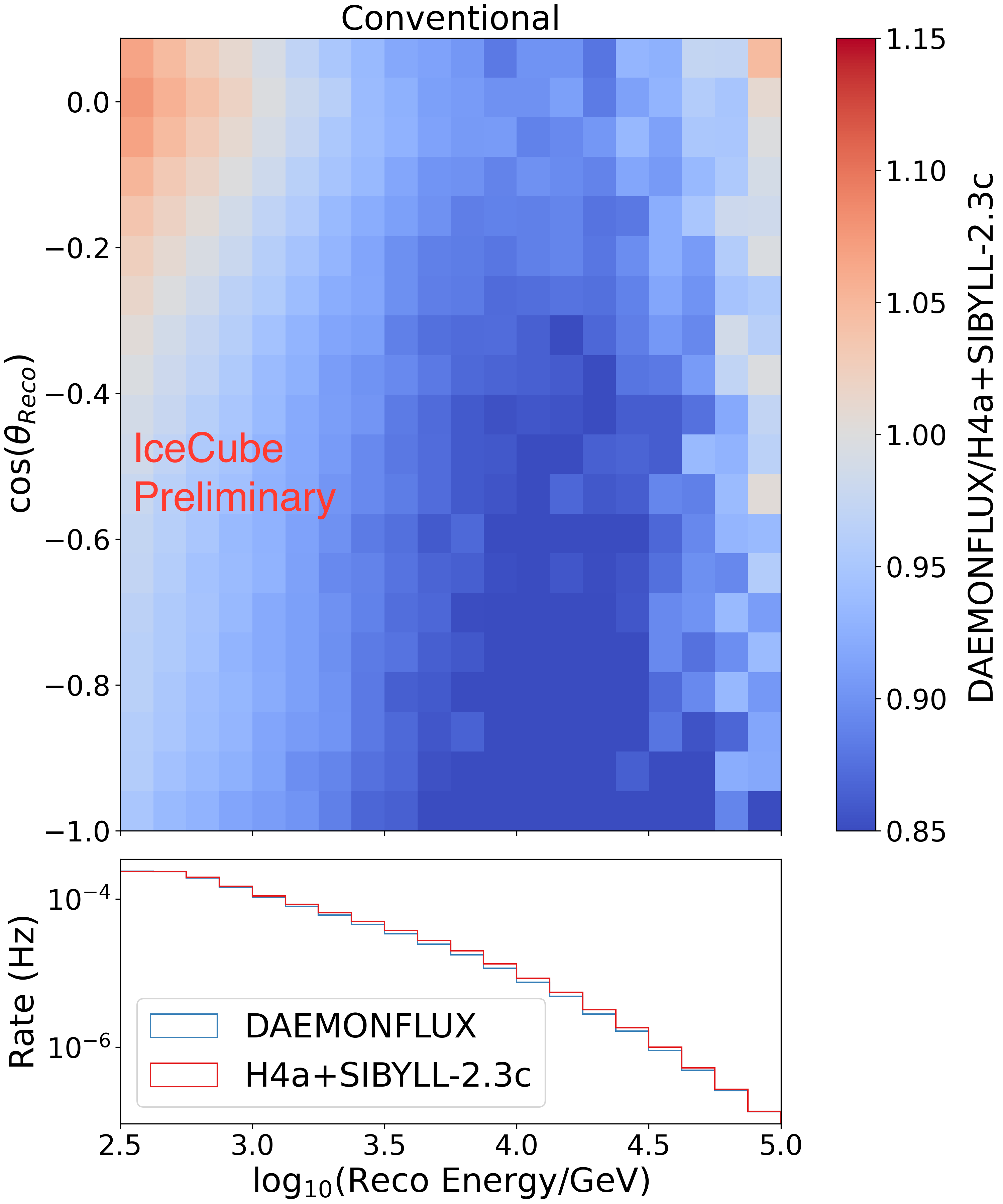}
    \end{minipage}
    \hspace{0.02\linewidth}
    \begin{minipage}{0.48\linewidth}
        \raggedright
        \includegraphics[width=0.97\linewidth]{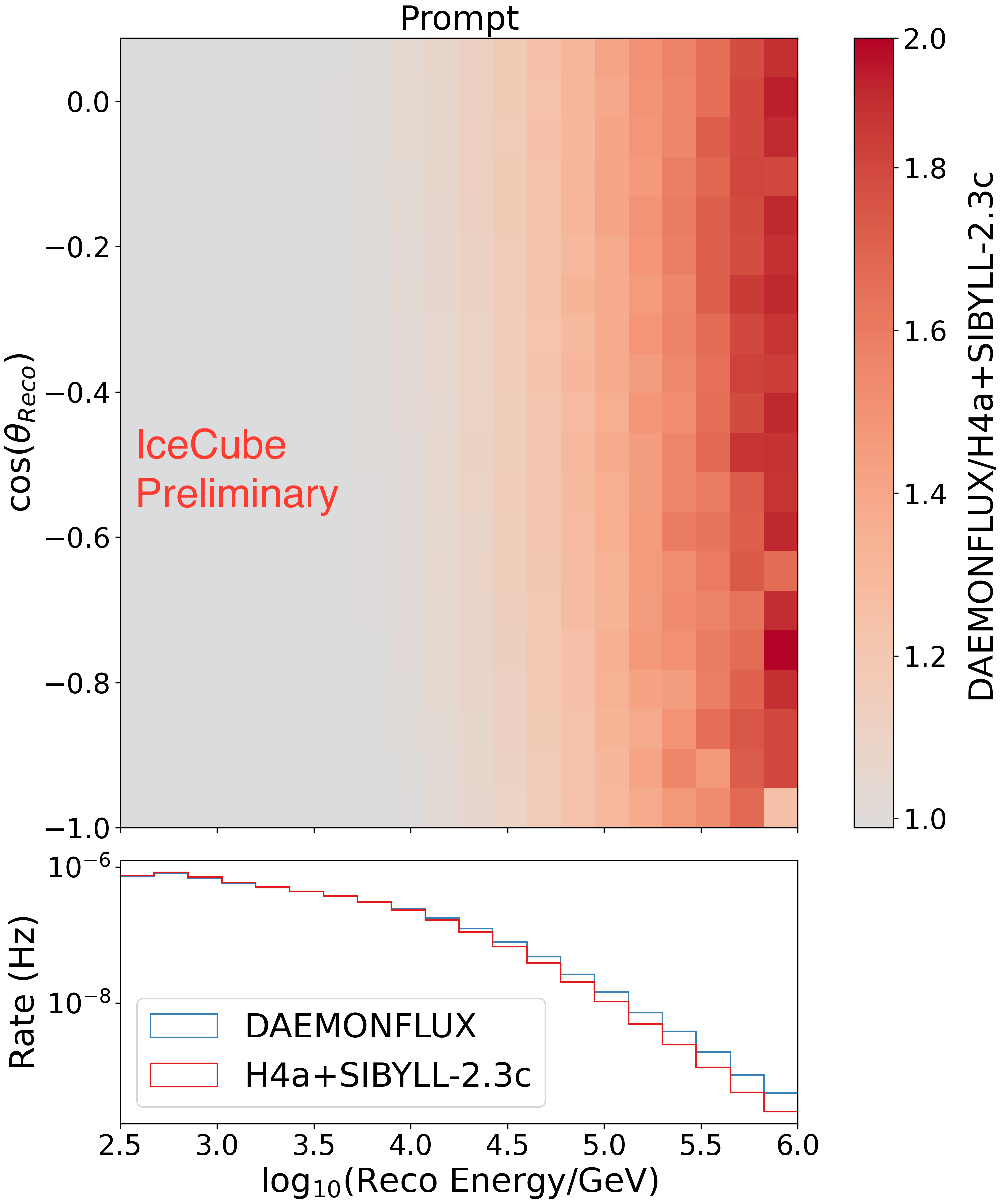}
    \end{minipage}
    \caption{Nominal conventional (left) and prompt (right) atmospheric neutrino flux comparison of DAEMONFLUX and H4a+SIBYLL-2.3c modeling with the atmospheric neutrino dominated throughgoing tracks sample.}
    \label{fig:df_to_barr}
\end{figure}

A comparison of the nominal conventional and prompt neutrino fluxes in the atmospheric neutrino–dominated throughgoing tracks sample between the two modeling schemes, as illustrated in Figure~\ref{fig:df_to_barr}, reveals less than a 15\% difference in the conventional flux, but a much larger difference—exceeding 100\% in the high-energy region—for the prompt flux.

Predictions of the prompt neutrino flux carry large uncertainties due to their strong dependence on hadronic interaction models, which suffer from limited experimental constraints as the relevant phase space is inaccessible to fixed-target experiments \cite{daemonflux}. The degeneracy between prompt and astrophysical neutrino fluxes, arising from their similar arrival zenith angle and energy distributions up to PeV energies, is stronger in the DAEMONFLUX scheme compared to the Barr scheme due to the higher predicted prompt flux at high energies.

To investigate the impact of this modeling update on sensitivity to our signal parameters, we injected an SPL astrophysical neutrino flux with best-fit values taken from \cite{Abbasi:diffuse_numu}, and performed profile likelihood scans for the astrophysical flux normalization. Alongside these, we also plotted the freely fitted normalizations for the atmospheric conventional and prompt flux components.

\begin{figure}[h]

\begin{subfigure}{0.5\textwidth}
\includegraphics[width=0.87\linewidth]{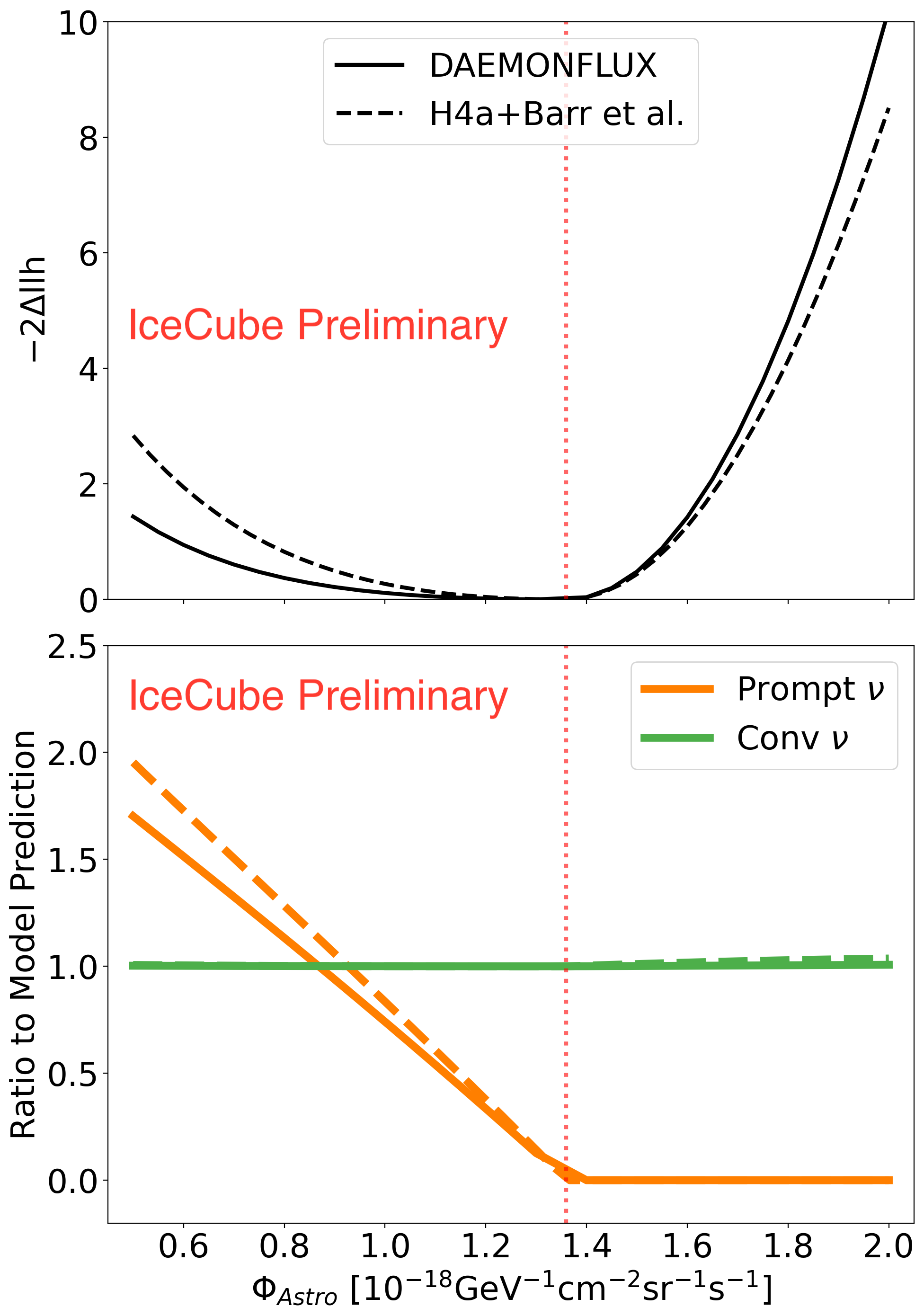} 
\caption{Throughgoing tracks}
\label{fig:astro_norm_scan}
\end{subfigure}
\begin{subfigure}{0.5\textwidth}
\includegraphics[width=0.9\linewidth]{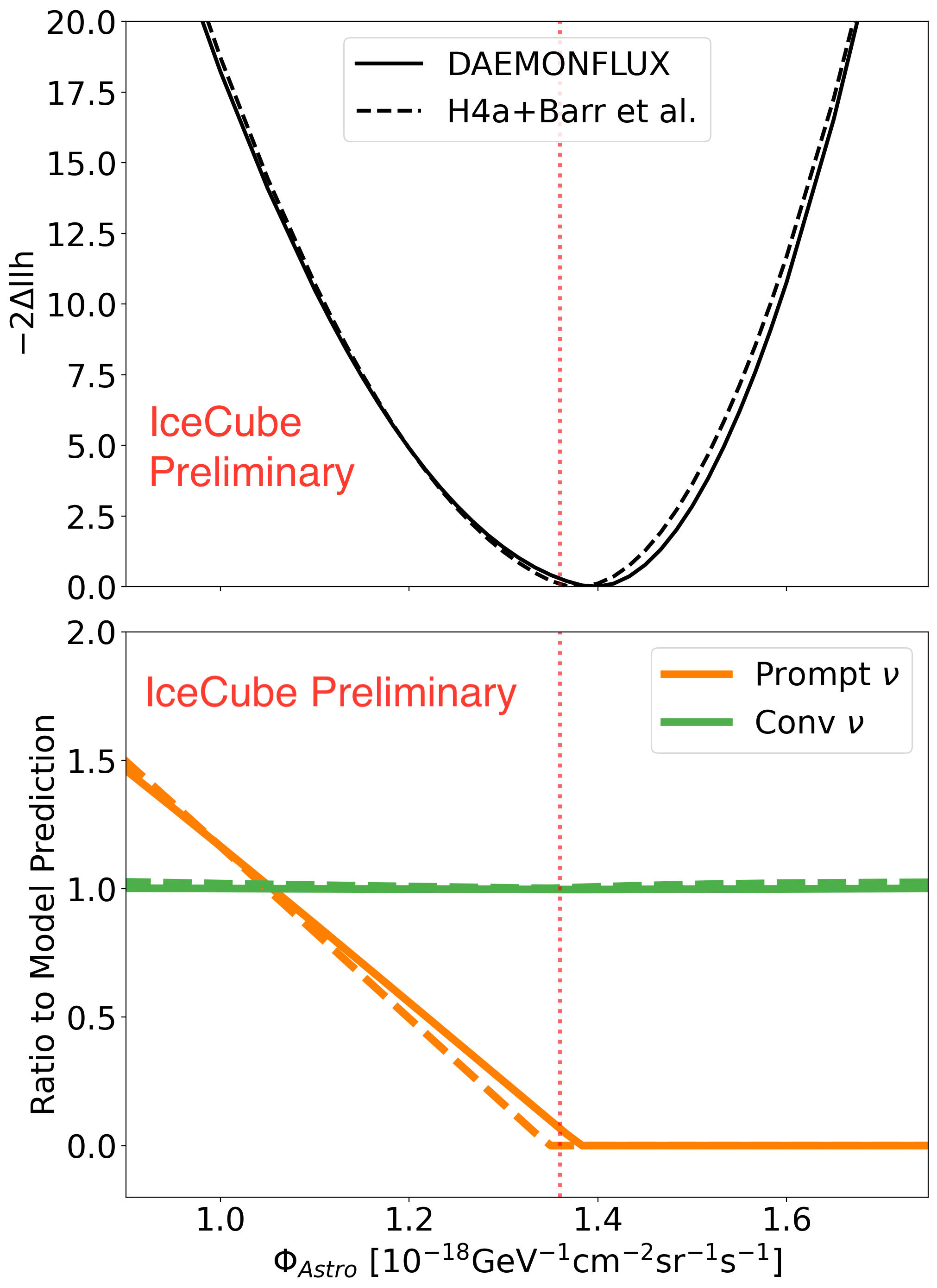}
\caption{Entire sample}
\label{fig:astro_norm_scan_2}
\end{subfigure}

\caption{1D likelihood scans of the astrophysical flux normalization using: (a) throughgoing tracks, and (b) the full combined sample. In the upper panels, solid (dashed) black curves show the likelihood profiles using the DAEMONFLUX (H4a+Barr et al.) atmospheric flux modeling, while the dotted red lines indicate the injected values. The lower panels show the ratios of the freely fitted atmospheric neutrino flux normalizations to their respective model predictions, with injected values of 1 for the conventional and 0 for the prompt.}
\label{fig:image2}
\vspace{-11pt}
\end{figure}

In an initial test using only the throughgoing tracks sample (Figure~\ref{fig:astro_norm_scan}), we observed that when the astrophysical normalization is fixed to values below the injected one, the prompt flux normalization gradually increases in both DAEMONFLUX and Barr schemes to compensate for the missing astrophysical component. As expected, the stronger degeneracy in the DAEMONFLUX scheme leads to a degraded sensitivity. Conversely, when the astrophysical normalization is fixed above the injected value, the already non-existent prompt contribution—consistent with previous IceCube results \cite{Abbasi:diffuse_numu,estes}—cannot account for the excess, rendering the degeneracy irrelevant and resulting in similar sensitivities for both schemes.

The second test, performed using the entire sample, shows that the degradation in sensitivity is mitigated in the combined likelihood fit, as illustrated in Figure~\ref{fig:astro_norm_scan_2}, with the degeneracy broken in the Southern sky by the self-veto effect.

\subsection{Atmospheric muon treatment}\label{subsec5}

The overwhelming atmospheric muon background is suppressed to levels orders of magnitude below the neutrino flux using a combination of traditional veto-based and neural network–based methods for each sample. However, two related challenges persist even after this suppression: the dependence of muon production on hadronic interaction models for both conventional and prompt fluxes, and the limited Monte Carlo simulation statistics. To improve our coverage of production uncertainties, we include the prompt muon flux for the first time in a full-sky astrophysical diffuse analysis, alongside the conventional muon flux, with their normalizations determined in a data-driven manner due to the lack of experimental and theoretical constraints. To address the low Monte Carlo statistics, we construct templates for atmospheric muon fluxes in each sample using a kernel density estimation (KDE) method.

\section{Sensitivity to substructures in PeV energies}\label{sec4}
\begin{figure}[h]
\vspace{-7.5pt}
\centering
\includegraphics[width=0.8\linewidth]{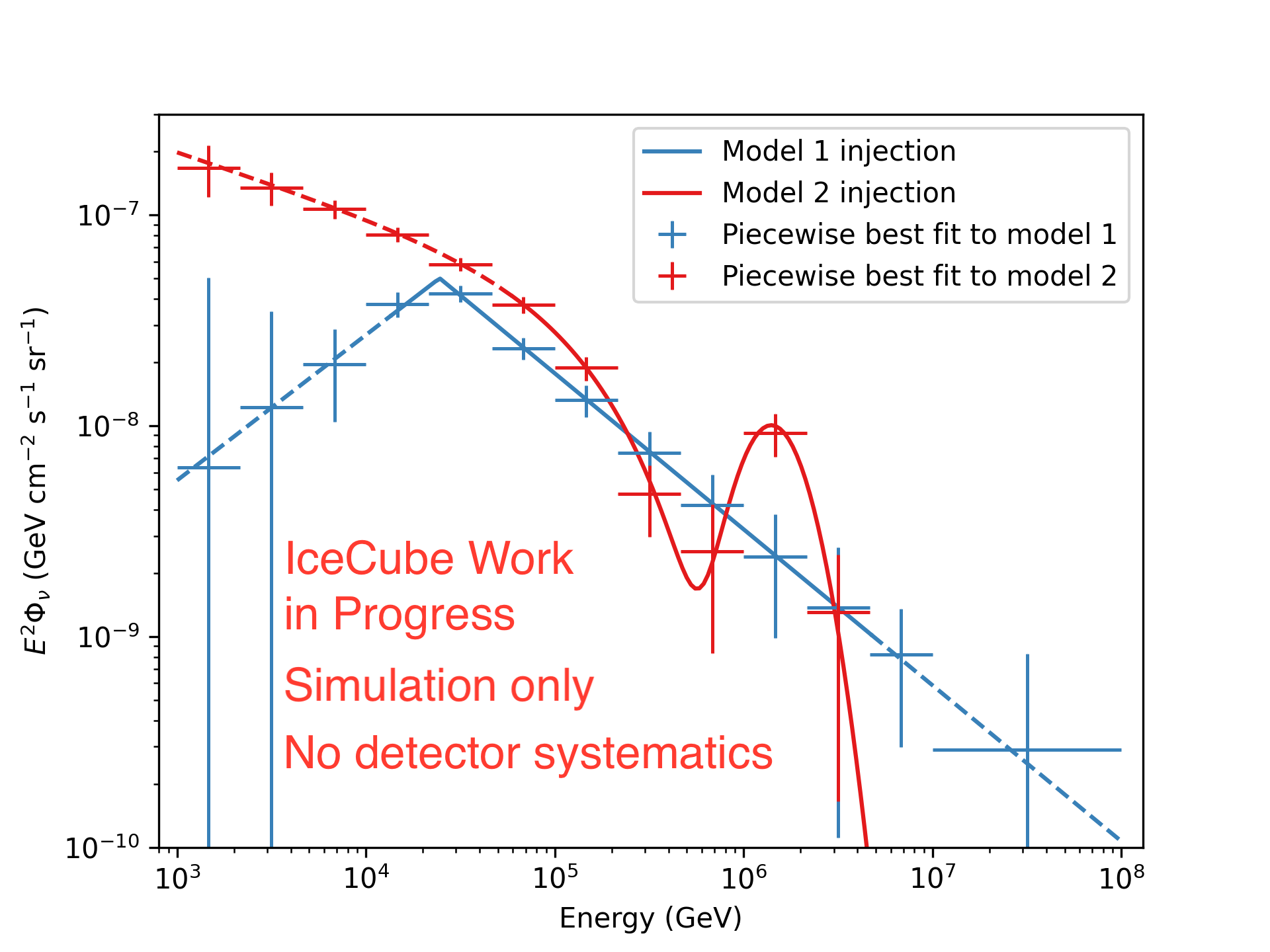}
\caption{Expected differential neutrino fluxes under two flux scenarios. Blue and red solid lines show the BPL and the SPL with a cutoff plus bump models from \cite{Abbasi:globalfit-2023} and \cite{bump-hunting}, respectively. Blue (red) dashed lines indicate the extrapolation of the BPL (SPL with cutoff plus bump) model beyond its reported sensitive energy range (binning range). Blue (red) crosses represent the simulation-only best-fit differential fluxes from this combined sample when fitting to the injected BPL (SPL with cutoff plus bump) model. The 1$\sigma$ error bars are derived from the Hessian matrix, incorporating statistical and atmospheric systematics uncertainties.}\label{fig:piecewise_1}
\vspace{-15pt}
\end{figure}

To test our sample’s capability to resolve features at PeV energies, we evaluated two astrophysical flux scenarios based on the recent IceCube result from a combined cascade and track analysis \cite{Abbasi:globalfit-2023} and the best-fit to public high-energy IceCube data \cite{bump-hunting}. The first model, a broken power law (BPL), predicts a soft spectrum above a break energy, 24.5 TeV, and a harder spectrum below it. The second model, an SPL with a cutoff and bump (Eq.~\ref{eq:model2}), predicts a substructure at PeV energies. Figure~\ref{fig:piecewise_1} illustrates the sample’s potential to probe features in the PeV region.

\begin{equation}
\begin{aligned}
\Phi_{\nu} =\Phi_{0,\mathrm{pl}} \left( \frac{E_\nu}{100 \mathrm{TeV}} \right)^{2-\gamma} e^{-\frac{E_\nu}{E_{\nu,\mathrm{cutoff}}}}+ \Phi_{0,\mathrm{bump}} \left(\frac{E_{\nu,\mathrm{bump}}}{E_\nu} \right)^2 e^{
\left[ - \alpha_\mathrm{bump} \mathrm{log}^2 \left( \frac{E_\nu}{E_{\nu,\mathrm{bump}}}
\right)
\right]}.
\label{eq:model2}
\end{aligned}
\end{equation}

\section{Conclusion}\label{sec5}
Substructure in the neutrino spectrum beyond the SPL was first reported by IceCube through a combined fit of multiple detection channels with a consistent treatment of systematic uncertainties \cite{Abbasi:globalfit-2023}. In this work, we present an updated combined analysis incorporating additional detection channels, improved event reconstructions, and enhanced systematic modeling. Preliminary results indicate sensitivity capable of probing the PeV energy range.

\bibliographystyle{ICRC}
\bibliography{references}

%

\clearpage

\input{authorlist_IceCube.tex}

\end{document}

%% file: ICRCdetails.tex
\ConferenceLogo{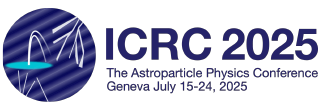}

\FullConference{39th International Cosmic Ray Conference (ICRC2025)\\
 15–24 July 2025\\
Geneva, Switzerland\\}

%% file: authorlist_IceCube.tex
\section*{Full Author List: IceCube Collaboration}

\scriptsize
\noindent
R. Abbasi$^{16}$,
M. Ackermann$^{63}$,
J. Adams$^{17}$,
S. K. Agarwalla$^{39,\: {\rm a}}$,
J. A. Aguilar$^{10}$,
M. Ahlers$^{21}$,
J.M. Alameddine$^{22}$,
S. Ali$^{35}$,
N. M. Amin$^{43}$,
K. Andeen$^{41}$
% ...
\\
\\
% ...
$^{10}$ Universit{\'e} Libre de Bruxelles, Science Faculty CP230, B-1050 Brussels, Belgium \\
% ...
$^{16}$ Department of Physics, Loyola University Chicago, Chicago, IL 60660, USA \\
$^{17}$ Dept. of Physics and Astronomy, University of Canterbury, Private Bag 4800, Christchurch, New Zealand \\
% ...
$^{21}$ Niels Bohr Institute, University of Copenhagen, DK-2100 Copenhagen, Denmark \\
$^{22}$ Dept. of Physics, TU Dortmund University, D-44221 Dortmund, Germany \\
% ...
$^{35}$ Dept. of Physics and Astronomy, University of Kansas, Lawrence, KS 66045, USA \\
% ...
$^{39}$ Dept. of Physics and Wisconsin IceCube Particle Astrophysics Center, University of Wisconsin{\textemdash}Madison, Madison, WI 53706, USA \\
% ...
$^{41}$ Department of Physics, Marquette University, Milwaukee, WI 53201, USA \\
% ...
$^{43}$ Bartol Research Institute and Dept. of Physics and Astronomy, University of Delaware, Newark, DE 19716, USA \\
% ...
$^{63}$ Deutsches Elektronen-Synchrotron DESY, Platanenallee 6, D-15738 Zeuthen, Germany \\
$^{\rm a}$ also at Institute of Physics, Sachivalaya Marg, Sainik School Post, Bhubaneswar 751005, India \\
% ...

\subsection*{Acknowledgments}

\noindent
The authors gratefully acknowledge the support from the following agencies and institutions:
USA {\textendash} U.S. National Science Foundation-Office of Polar Programs,
U.S. National Science Foundation-Physics Division,
U.S. National Science Foundation-EPSCoR,
U.S. National Science Foundation-Office of Advanced Cyberinfrastructure,
Wisconsin Alumni Research Foundation,
Center for High Throughput Computing (CHTC) at the University of Wisconsin{\textendash}Madison,
Open Science Grid (OSG),
Partnership to Advance Throughput Computing (PATh),
Advanced Cyberinfrastructure Coordination Ecosystem: Services {\&} Support (ACCESS),
Frontera and Ranch computing project at the Texas Advanced Computing Center,
U.S. Department of Energy-National Energy Research Scientific Computing Center,
Particle astrophysics research computing center at the University of Maryland,
Institute for Cyber-Enabled Research at Michigan State University,
Astroparticle physics computational facility at Marquette University,
NVIDIA Corporation,
and Google Cloud Platform;
Belgium {\textendash} Funds for Scientific Research (FRS-FNRS and FWO),
FWO Odysseus and Big Science programmes,
and Belgian Federal Science Policy Office (Belspo);
Germany {\textendash} Bundesministerium f{\"u}r Forschung, Technologie und Raumfahrt (BMFTR),
Deutsche Forschungsgemeinschaft (DFG),
Helmholtz Alliance for Astroparticle Physics (HAP),
Initiative and Networking Fund of the Helmholtz Association,
Deutsches Elektronen Synchrotron (DESY),
and High Performance Computing cluster of the RWTH Aachen;
Sweden {\textendash} Swedish Research Council,
Swedish Polar Research Secretariat,
Swedish National Infrastructure for Computing (SNIC),
and Knut and Alice Wallenberg Foundation;
European Union {\textendash} EGI Advanced Computing for research;
Australia {\textendash} Australian Research Council;
Canada {\textendash} Natural Sciences and Engineering Research Council of Canada,
Calcul Qu{\'e}bec, Compute Ontario, Canada Foundation for Innovation, WestGrid, and Digital Research Alliance of Canada;
Denmark {\textendash} Villum Fonden, Carlsberg Foundation, and European Commission;
New Zealand {\textendash} Marsden Fund;
Japan {\textendash} Japan Society for Promotion of Science (JSPS)
and Institute for Global Prominent Research (IGPR) of Chiba University;
Korea {\textendash} National Research Foundation of Korea (NRF);
Switzerland {\textendash} Swiss National Science Foundation (SNSF).